\documentclass[a4paper]{article}
\usepackage[english]{babel}
\usepackage[utf8]{inputenc}
\usepackage[T1]{fontenc}

\usepackage[a4paper,top=3cm,bottom=2cm,left=3cm,right=3cm,marginparwidth=1.75cm]{geometry}

\usepackage{amsmath, amsthm, amsfonts}
\usepackage{graphicx}
\usepackage[colorlinks=true, allcolors=blue]{hyperref}
\usepackage{url}
\title{Intrinsic time resolution and efficiency characterization for two size of plastic scintillators studies in Geant4}
\author{\normalsize \large \bf 
\thanks{Corresponding author: hzepeda@fcfm.buap.mx} C.~H.~Zepeda~Fern\'andez$^{1,2}$, 
  L. F. Rebolledo-Herrera$^2$,\\
\normalsize \large \bf E.~Moreno~Barbosa$^2$, E.~M\'arquez~Quintos$^2$\\
 \normalsize \small \it{$^1$Cátedra CONACyT, 03940, CdMx Mexico}\\
\normalsize \small \it{$^2$Facultad de Ciencias F\'isico Matem\'aticas,} \\
\normalsize \small \it{Benem\'erita Universidad Aut\'onoma de Puebla,}\\
\normalsize \small \it{Av. San Claudio y 18 Sur, Ciudad Universitaria 72570, Puebla, Mexico}
}

\begin{document}
\maketitle

\begin{abstract}
The time resolution (TR) is one of the most important characteristic of a detector. For the case of  a scintillator, the collection of light also is important, it depends of the effective area of the photo-sensor (\textit{Scorer}). The Photomultipliers Tubes (PMTs), normally, have an effective area greater than the Silicon Photomultipliers (SiPMs), however, due to its voltage of operation, their size and  cost, in some cases it can be difficult to work with them. In which case, it is preferable to use SiPMs. The value of TR also depends of the size and geometry of the scintillator, number of photo-sensors and the electronic part.\\  
In this work, we study the mean optical photon arrival time distribution (AT) to a Scorer from a SiPM of 6$\times$6~mm$^2$. We define the variation of AT as \textit{the intrinsic time resolution} (ITR). In Geant4, we simulated two different geometries: square and hexagonal, for a BC-404 plastic scintillator coupled to one Scorer. The sources simulated were Sr$^{90}$, Co$^{60}$, Cs$^{137}$, Na$^{22}$ and $\mu^-$ of 1~GeV. It is shown that AT and ITR  depends of the geometry and size of the plastic scintillator, the location of the Scorer, the incident particle and its energy. Then, the ITR and therefore the TR is not a constant for a detector. Finally, we show the relation between AT and the deposited energy by the particle incident, which are related in the experiment to the response time event of the detector and the deposited charge by the incident particle, respectively. 
\end{abstract}

\section{Introduction}
One of the most important characteristic for a detector is the time resolution (TR), many works have been done to improve it~\cite{ALVARADO2020163150}. For the case of the scintillator materials, the detection of optical photons is very important to give the signal, of which, photo-sensors are used, then, the  detection area  (\textit{Scorer}) of them is important, if it is bigger, it has better light collection. Two kind of photo-sensors are commonly used: the Photon Multiplier Tube (PMT) and Silicon Photon Multiplier (SiPM). The PMTs have larger dimensions  than SiPMs, which have a size of the order of cm, while SiPMs have a size of mm. Therefore, geometrically  PMTs have a larger Scorer than  SiPMs. However, when it is required to work with a small plastic scintillator (mm$^3$) using a SiPM is more convenient, due to its Scorer~\cite{Kado_2021}. Recently, the use of SiPMs have been used more than PMTs by getting a better response and signal~\cite{Garutti,Simon,Kuper}. A 6$\times$6~mm$^3$ SensL SiPM has two kind of output signals: the \textit{standard} and \textit{fast}~\cite{sensl}. The fast signal pulse has a rise time of 1~ns and a pulse width of 3.2~ns, while, the rise time of the standard signal pulse is 4~ns and a width of 100~ns. From the standard pulse, it is possible to reconstruct the deposited charge, making an integration of it, for this reason, reading this pulse gives physical information. However, recently, it has been shown that the deposited charge can be reconstructed from the fast   signal~\cite{Zepeda_Fern_ndez_2020}. Therefore, this signal can be taken for a higher detector response, due to its rise time and improve the TR value. \\
Usually crystals or plastics are used in scintillation detectors, depending on the particles to be detected, it is chosen which one to use.  Recently, with the aim to improve the time and spatial resolution, the  coupling between SiPM and scintillator material has been widely investigated 
~\cite{Ebru,Wieczorek}. The plastic scintillators have been  more frequently used
~\cite{Krzemien,Raczynski} due to its low cost compared with crystals. In particular, the BC404 plastic scintillator is being used in the construction of a new Mexican detector~\cite{Kado_2021} within the Multi Purpose Detector experiment tat the Nuclotron Ion Collider fAcility (NICA) ~\cite{MPD} .\\
The use of detector simulations is an important study, mainly to have a control on the physical parameters. GEANT4 is one of many software which allows to simulate them~\cite{GHADIRI201563,Usubov:2013yma} and help the experimental measurements. As it was mentioned above, the use of SiPMs has shown some advantages when reading the signal and, simulations in this field have been developed~\cite{Leming_2014,Xi,Cates_2018,OTTE2006417}.\\
We define the \textit{intrinsic time resolution} (ITR) by the standard deviation of the \textit{mean of the optical photon arrival time distribution} (AT) to the Scorer. Then, the ITR is an intrinsic property which depends of the geometry, scintillator and Scorer sizes, but also the scorer location attached to the scintillator. The number of Scorers also modifies  the value of ITR, in particular for its improvement~\cite{ALVARADO2020163150}. In the experiment, the TR ($\sigma_{TR}$) is measured and its relation with the ITR ($\sigma_{ITR}$) is,\\
\begin{equation}\label{Exp-Sim}
   \displaystyle  \sigma_{TR}^2 =\sigma_{ITR}^2+\sigma_{ele}^2+\sigma_{ref}^2.
\end{equation}
Where $\sigma_{ele}$ and $\sigma_{ref}$ are the time resolution of the electronics and the reference trigger detector, respectively~\cite{ALVARADO2020163150}.\\
In this work, we show the value of ITR and AT as function of the BC-404 plastic  scintillator size and the location of the Scorer, as well as the incident particles and their energy. 
Finally, we show an estimate result in the experimental measurements. 

\section{Simulation methodology}
The Geant4 10.7~\cite{geant4} was used for this simulation.  All optical properties for the BC-404 plastic scintillator were considered as it is described below.  

\subsection{Plastic scintillator BC-404 simulation}
\begin{figure}[htbp]
\begin{center}
\includegraphics[width=0.8\textwidth]{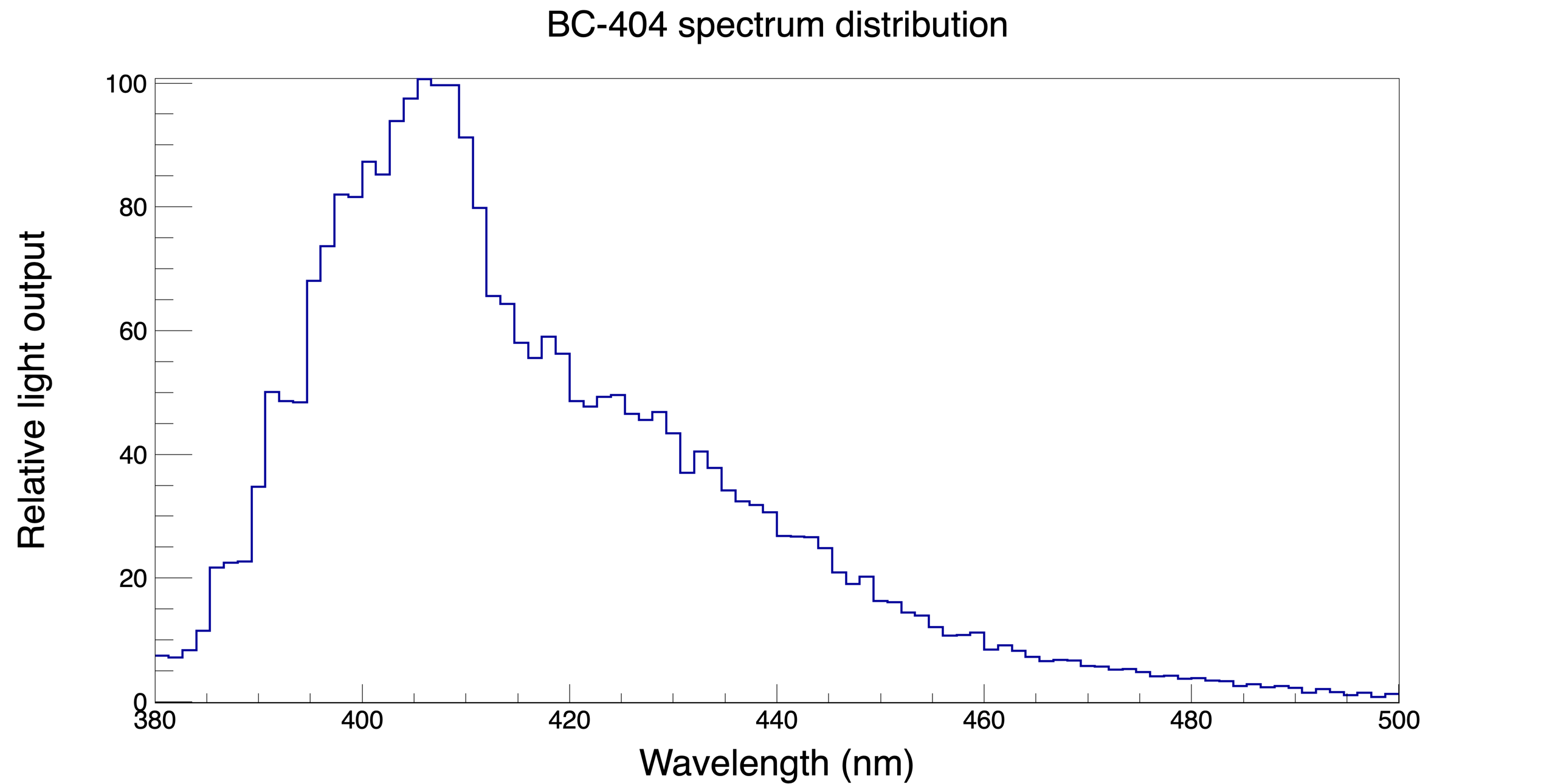}
\end{center}
\caption{BC-404 plastic scintillator emission spectra simulation.}
\label{spectra}
\end{figure}

The polyvinyltoluene material for BC-404 plastic scintillator is already defined in the environment of Geant4 by \textit{G4$\_$PLASTIC$\_$SC$\_$VINYLTOLUENE}.  All the optical properties used for the simulation have been taken from the Saint Gobain BC-404 data sheet~\cite{bc404data} and the Light output property was taken from a thesis study~\cite{phdthesis}. Some of these optical  properties are shown in Table~\ref{bc404}. The emission spectra simulated based is shown in Figure~\ref{spectra}.\\

\begin{table}[htbp]
\caption{Optical properties for the simulation of BC-404 plastic scintillator.}\label{bc404}
\centering
\smallskip
\begin{tabular}{|c |c |c|c|}
\hline
Light output (photons /MeV) & Refraction index & Light attenuation length (m)& Wavelength of\\
& & & Max. Emission (nm)\\ 
\hline
10,880 & 1.58 & 1.4& 408\\
\hline
\end{tabular}
\end{table}

\subsection{Optical photons simulations}
The Geant4 environment models major optical physics in detector. It includes process as ionization, Compton, photoelectric, scintillation, Cerenkov and the resulting optical photon propagation. The main information for this work is from the optical photons, which are counted by the hits on the Scorer surface (see~\ref{conf}).
\subsection{Sources simulation}
Five radiation sources were simulated: Na$^{22}$, Cs$^{137}$, Co$^{60}$, Sr$^{90}$ and $\mu^-$. The first four sources belong to a kit that can be found in laboratories~\cite{kitsources} and for this study, it is considered their main branching ratio decay \cite{na22,cs137,co60,sr90}. The $\mu^-$ particles are considered as a cosmic ray average.  Based on these specifications the particles and energy for each source are shown in Table~\ref{source}. For the next sections, we refer as \textit{source} for the kit sources and \textit{$\mu^-$ source} for the muon. For each source 6,000 events were considered, where a single event is understood as a particle from the source which interacts with the BC-404 scintillator. Finally, the particles were concentrated in a radius of 2~mm, considering the effective aperture of the kit source. 

\begin{table}[htb]
\caption{Particle and energy simulated for each source.}
\centering
\begin{tabular}{|c|c|c|}
\hline
 Source & Particle & Energy (MeV)\\
\hline \hline
Na$^{22}$ & $\gamma$ & 0.511 \\ 
\cline{2-3}
& $\gamma$ & 1.275\\ \cline{1-3}
Cs$^{137}$ &$\gamma$ &0.6617\\ \cline{1-3}
Co$^{60}$ &$\gamma$ &1.170\\ \cline{1-3}
Sr$^{90}$ &$e^-$ &0.546\\ \cline{1-3}
$\mu^-$ &$\mu^{-}$ & 1,000\\ \cline{1-3}
\end{tabular}
\label{source}
\end{table}

\subsection{Configuration simulation}\label{conf}
\begin{figure}[htbp]
\begin{center}
\includegraphics[width=0.6\textwidth]{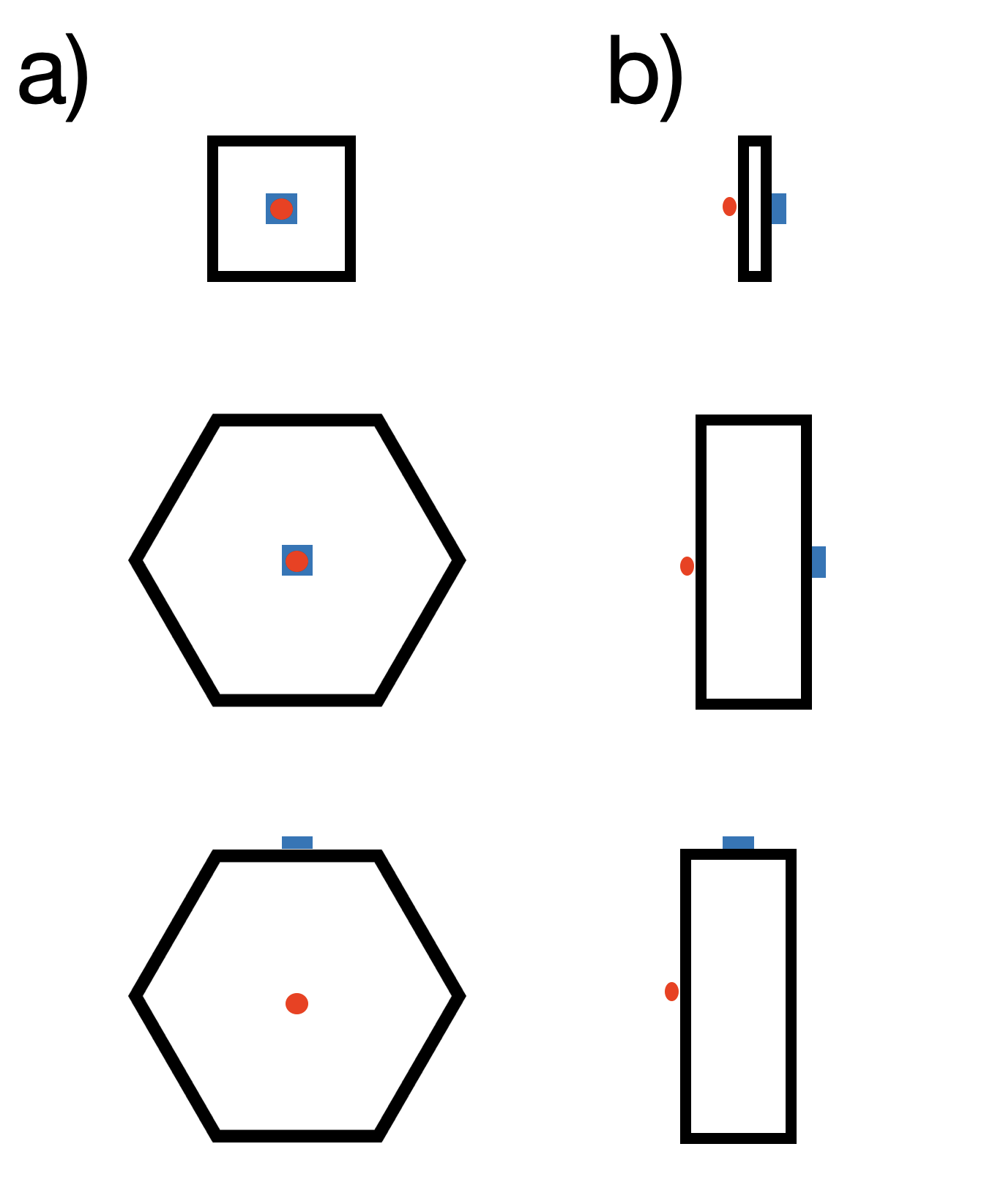}
\end{center}
\caption{Illustration for the three configurations. a) frontal and b) side view. The blue color square and hexagonal represents the BC-404 plastic scintillator, the black square represents the Scorer and the red circle represents the interaction point.}
\label{configurations}
\end{figure}

We simulated two geometries for the BC-404 plastic scintillator: 
\begin{itemize}
\item A hexagonal geometry with 50~mm high and 20~mm wide. For this geometry we considered two arrangements.   
\begin{itemize}
\item The Scorer coupled on one of the hexagonal faces at the geometrical center (first configuration).
\item The Scorer coupled at the center on one of the side faces (second configuration). 
\end{itemize}
\item A square geometry with a front face area of $20\times20~$mm$^2$ and a width of 3~mm (square configuration).
\end{itemize}
These configurations are shown in Figure~\ref{configurations}, the interaction point was simulated on the geometrical center of each geometry and located at 1~mm from the plastic scintillator. 
The Scorer was simulated with $6\times6$~mm$^2$, where the Scorer represents a SensL SiPM of this size as already mentioned. The configuration for the Scorer located on side face on the square configuration was not considered, due to its small width. The width of the Scorer is not relevant considering the reasons explained in~\ref{optical}.

Finally, all the optical photons physical parameters as energy, arrival position and arrival time were obtained on the Scorer surface. In addition, the deposited energy of the incident particle was saved and obtained in the BC-404 scintillator. This last quantity was used to obtain an estimation of the experimentral measurement (see~\ref{charge_estimation}).

\subsection{Optical boundaries}\label{optical}
The environment surrounding the plastic scintillator and the scorer was considered air by the material giving in Geant4: \textit{G4\_Air}. For this study are considered two boundary condition:
\begin{itemize}
    \item Scintillator-environment: It is considered with 95\% efficiency in reflection from the scintillator to environment surface.  The plastic scintillator surface was considered polished. Finally, the interface was considered by dielectric-metal. As we are not interested in the optical photons that come out of the scintillator, this condition is enough.
    \item Scintillator-Scorer: This surface was considered 100\% absorbent, in order that the optical photons that arrive to the Scorer, are not reflected and therefore do not count them more than once. Then, the interface is not relevant and it was considered by dielectric-metal, dielectric by scintillator and metal for the scorer. As the optical photons do not pass through the Scorer, its  width is not important.
\end{itemize}

\subsection{Intrinsic time resolution methodology}
To obtain the ITR of each configuration, we obtained the AT value, this value represents the average time at which the optical photons reach to the Scorer and it is obtained event by event. Finally, we obtained the fit  Gaussian distribution from the set of all mean values to obtain the $\sigma$ parameter, which it represents the ITR. This technique has been compared with the experiment~\cite{ALVARADO2020163150} and also is used to calculate the ITR for the  MiniBeBe detector (described in the Section 6 in~\cite{Kado_2021}).

\section{Results and analysis}
A plastic scintillator (in this case BC-404~\cite{bc404data}) is not particularly efficient for $\gamma$-particles detection in which case crystal scintillator is more often used, for example LYSO~\cite{lyso}. This effect was observed during simulation, where it was obtained that:
\begin{itemize}
    \item In the square geometry: for 50 $\gamma$-particles, in average, one of them interacts with the plastic scintillator.
    \item In the hexagonal geometry: for fifteen $\gamma$-particles, in average, two of them interacts with the plastic scintillator.
\end{itemize}
This difference in interaction is due to the width, for the square configuration the particles have a shorter interaction distance compared with the width of the hexagonal configurations.\\ 
For the $e^-$ and $\mu^-$ particles we obtained non-null events for all configurations.
\subsection{Square geometry}
\begin{figure}[htbp]
\begin{center}
\includegraphics[width=0.8\textwidth]{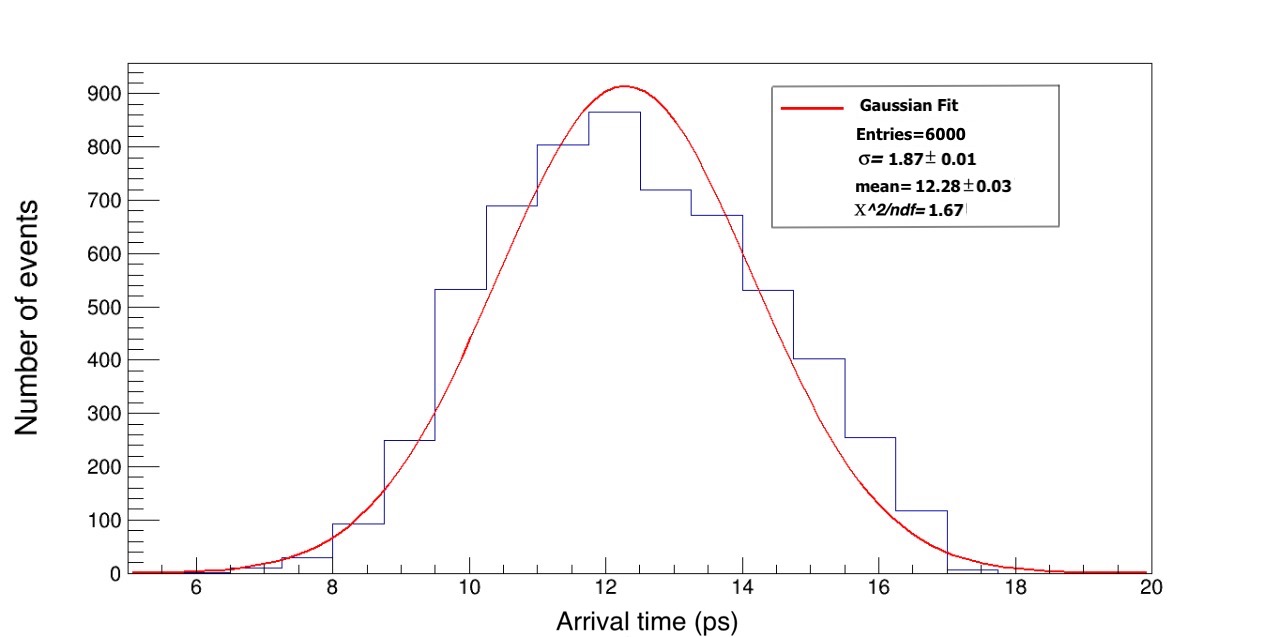}
\end{center}
\caption{The mean arrival time optical photon distribution for all events of $Sr^{90}$.}
\label{sr90minibebe}
\end{figure}

\begin{figure}[htbp]
\begin{center}
\includegraphics[width=0.8\textwidth]{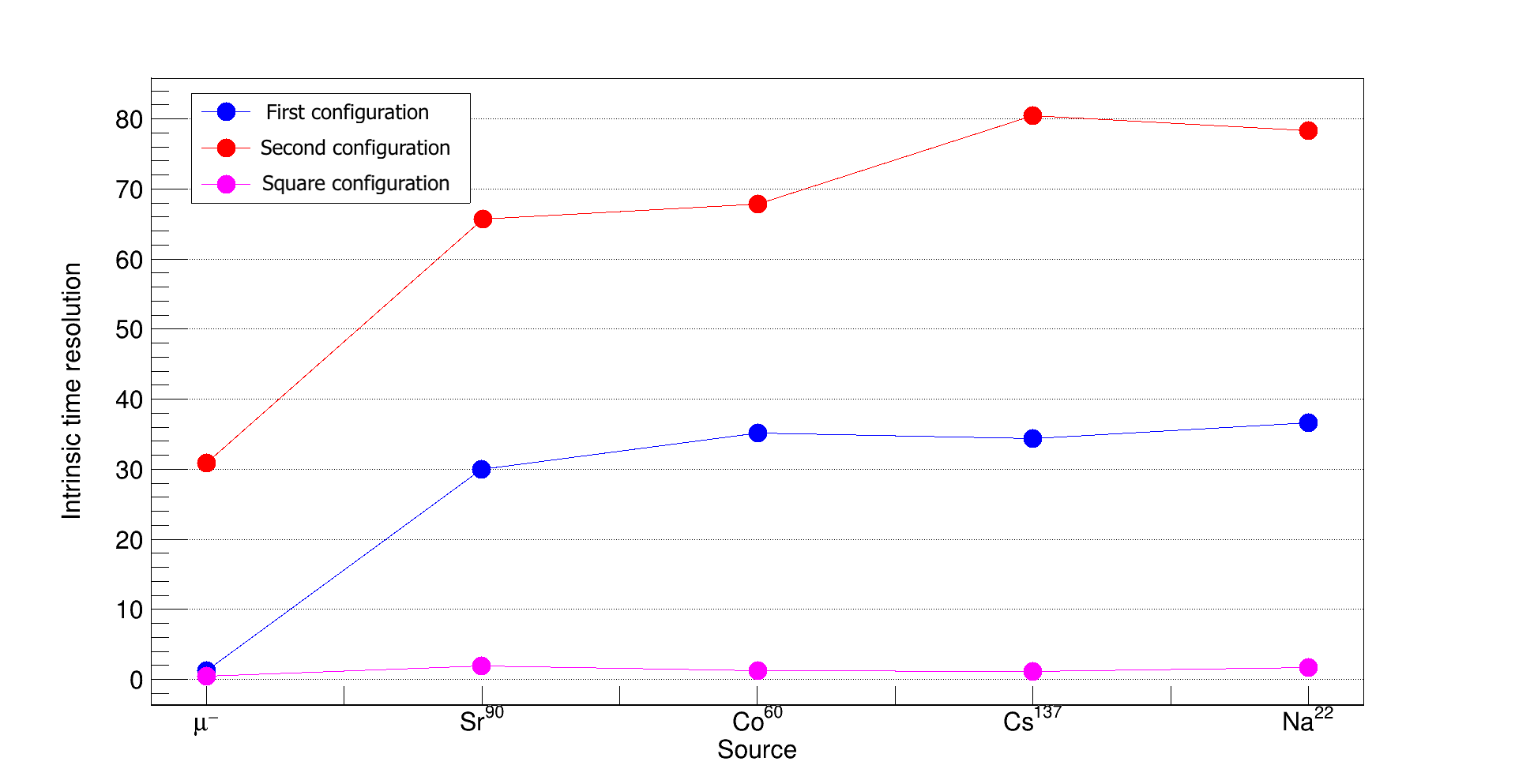}
\end{center}
\caption{The ITR values for all configurations.}
\label{ITR}
\end{figure}

\begin{figure}[htbp]
\begin{center}
\includegraphics[width=0.8\textwidth]{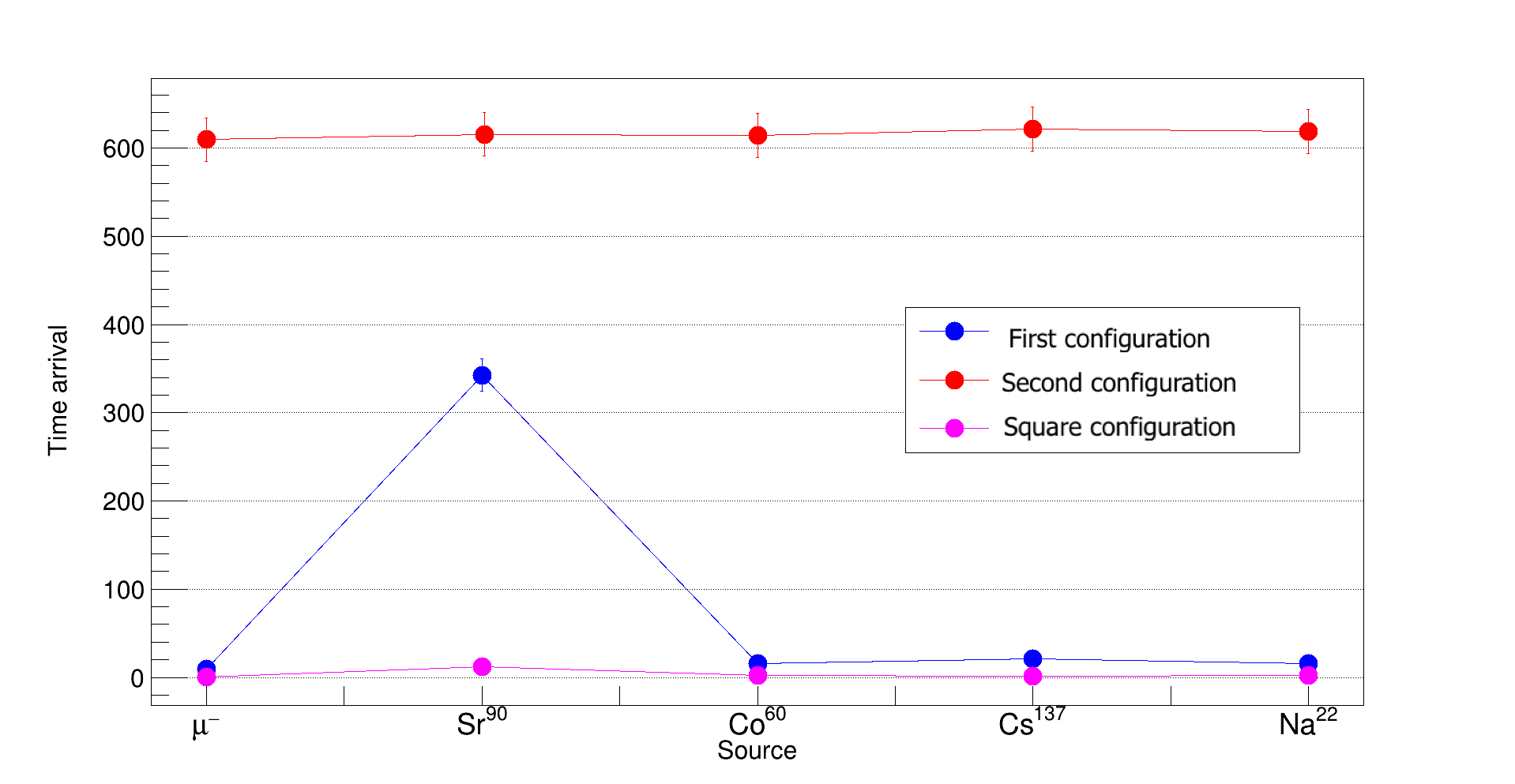}
\end{center}
\caption{The mean arrival time optical photon is shown for all sources.}
\label{mean}
\end{figure}

To exemplify the AT, in Figure~\ref{sr90minibebe} is shown for the case of $Sr^{90}$ source, where  the mean value of $12.28\pm0.03$~ps and an ITR value of $1.88\pm0.01$~ps were obtained. We obtained the respective distribution for all sources, the results are shown in Figures~\ref{ITR}~and~\ref{mean} (pink color). Due to the small size of the plastic, the ITR values for all sources are consistent around 2~ps.  The mean values (except for Sr$^{90}$) are consistent for all sources around 2~ps. These constant values are due to the small dimensions of the plastic scintillator, where the optical path is  small~\cite{TRCHZ}. The  mean value for the Sr$^{90}$ source is 12.26$\pm$0.02 ps, it will be discussed in the next subsection. 
For larger dimensions the ITR and AT  change as it is shown below.

\subsection{Hexagonal geometry}
\begin{figure}[htbp]
\begin{center}
\includegraphics[width=0.68\textwidth]{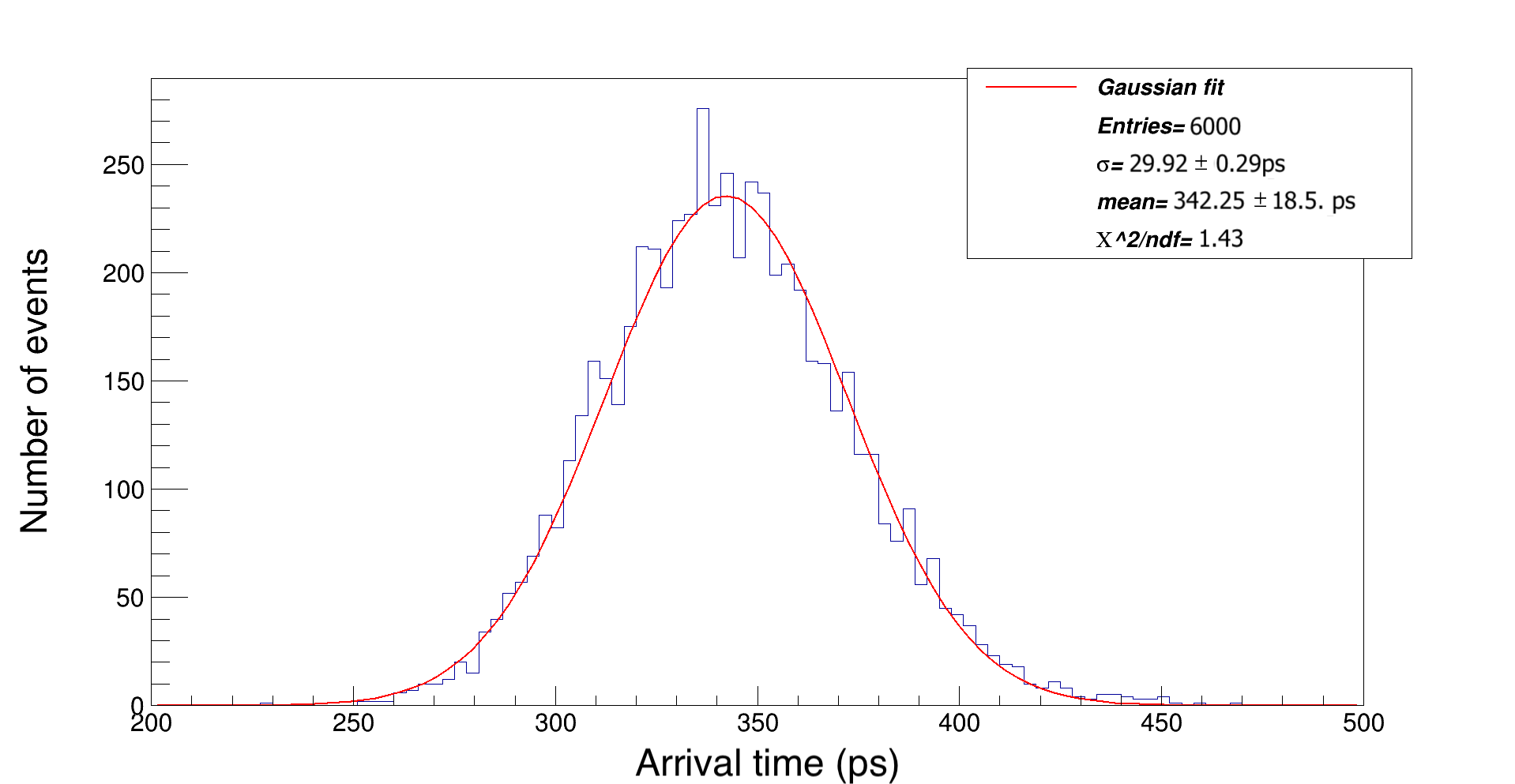}
\includegraphics[width=0.68\textwidth]{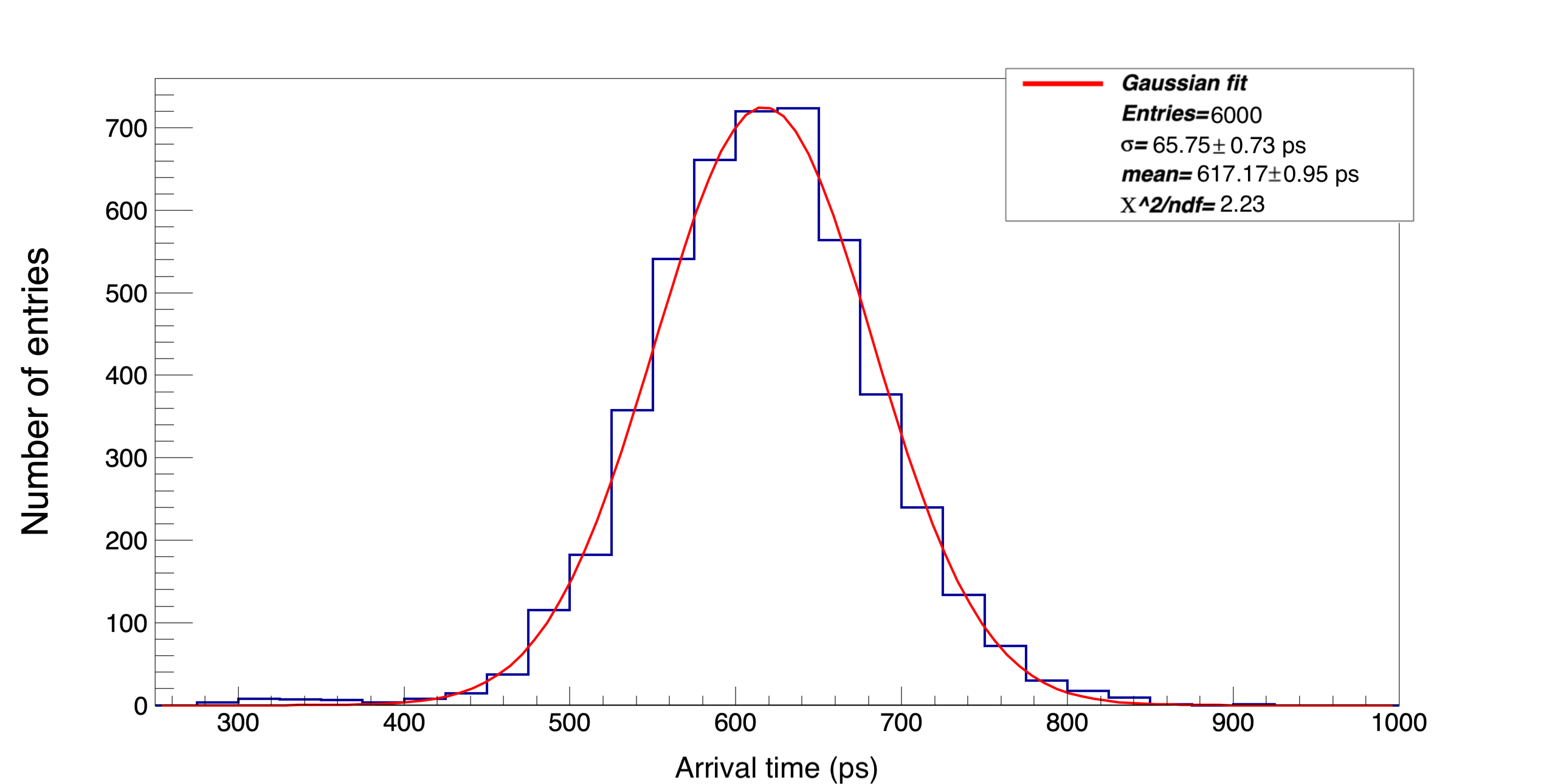}
\end{center}
\caption{The mean arrival time optical photon distribution for $Sr^{90}$ source for the Scorer located at center (up) and located at top (down).}
\label{Sr90TR}
\end{figure}

Once again to exemplify the AT, in Figure~\ref{Sr90TR} are shown for $Sr^{90}$ and both hexagonal configurations, getting an ITR value of $\sigma=29.92\pm0.29$~ps and $\sigma=65.75\pm0.73$~ps for the first and second configuration, respectively. The same results are obtained for the rest of the sources, as it is shown in Figure~\ref{ITR}. It can be seen that the ITR is greater for the second configuration  than the first  configuration for all sources. This is due to the mean  arrival time, resulting that the optical path to arrive to the Scorer is greater for the second configuration than the first configuration, as it is shown in Figure~\ref{mean}, it can be observed that for the second  configuration the optical photons take longer time to arrive at Scorer than  the first configuration, therefore the $\sigma$ value also is greater. The case of $\mu^-$ source has the lowest values for mean and $\sigma$, this is due to its great energy, by which more optical photons are created.
 
\begin{figure}[htbp]
\begin{center}
\includegraphics[width=0.8\textwidth]{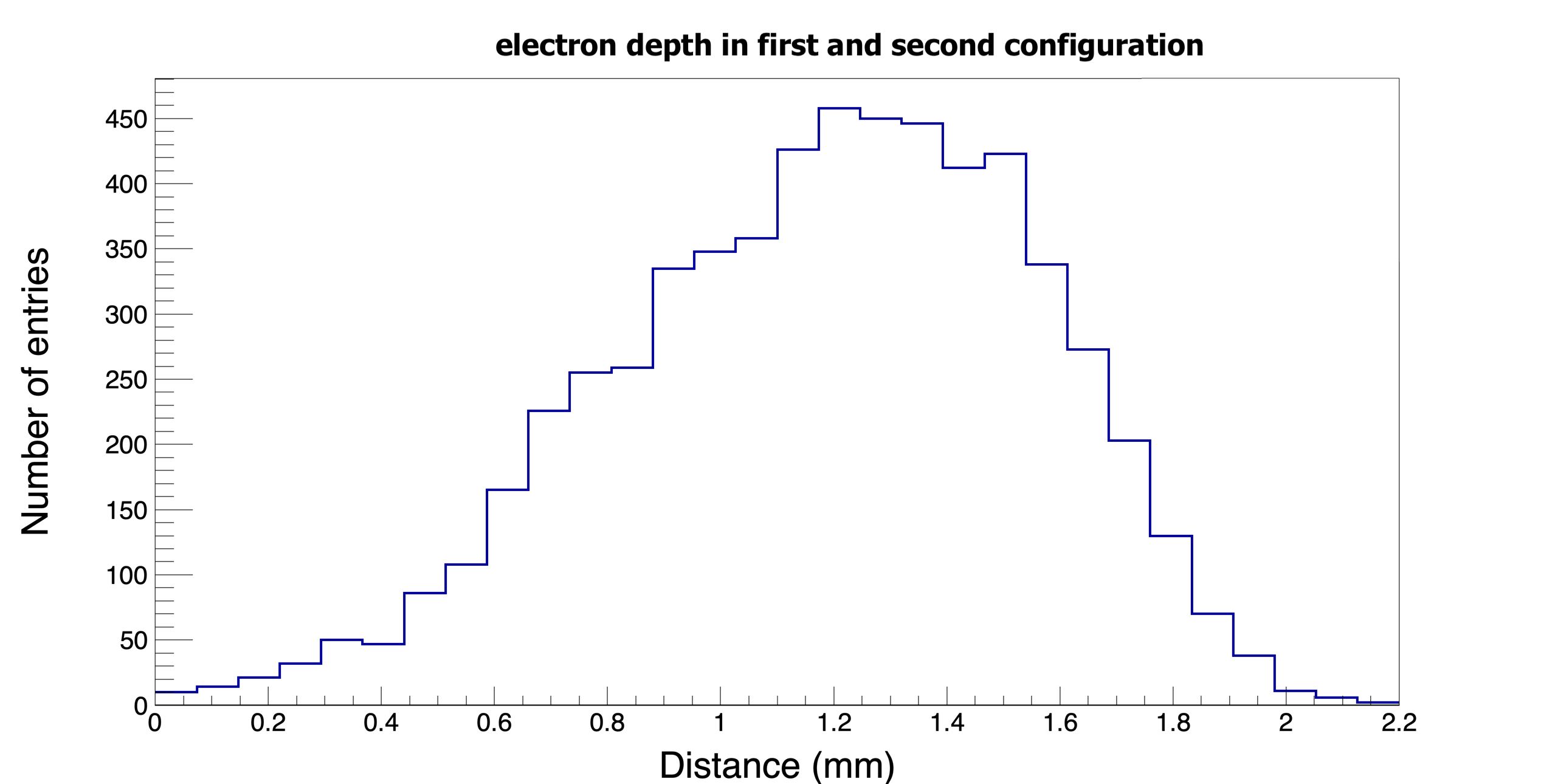}
\includegraphics[width=0.8\textwidth]{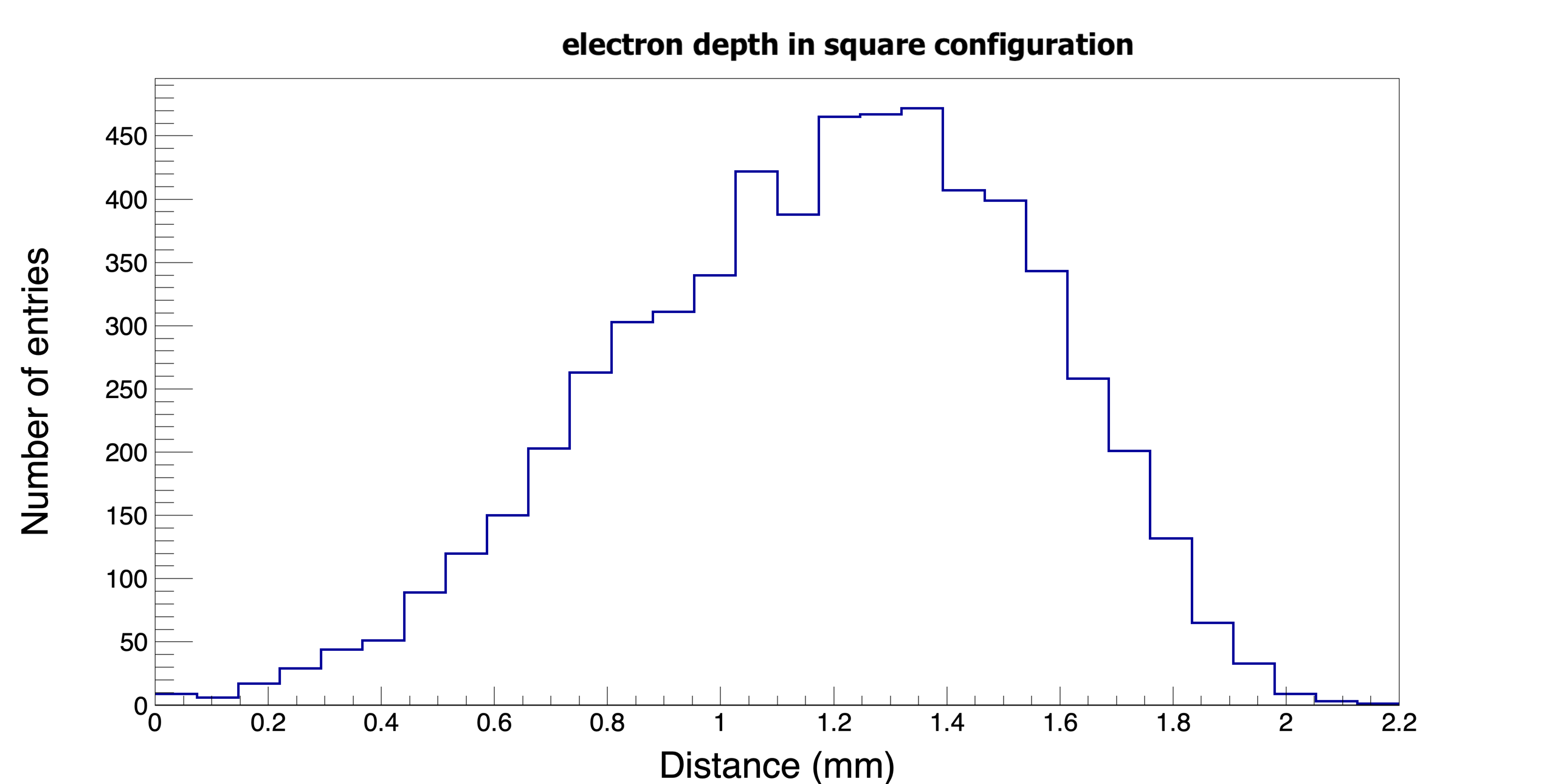}
\end{center}
\caption{The depth reached by the e$^-$ emited by Sr$^{90}$ source in the hexagonal (top) and square (down) geometry.}
\label{distance}
\end{figure}

\begin{figure}[htbp]
\begin{center}
\includegraphics[width=0.8\textwidth]{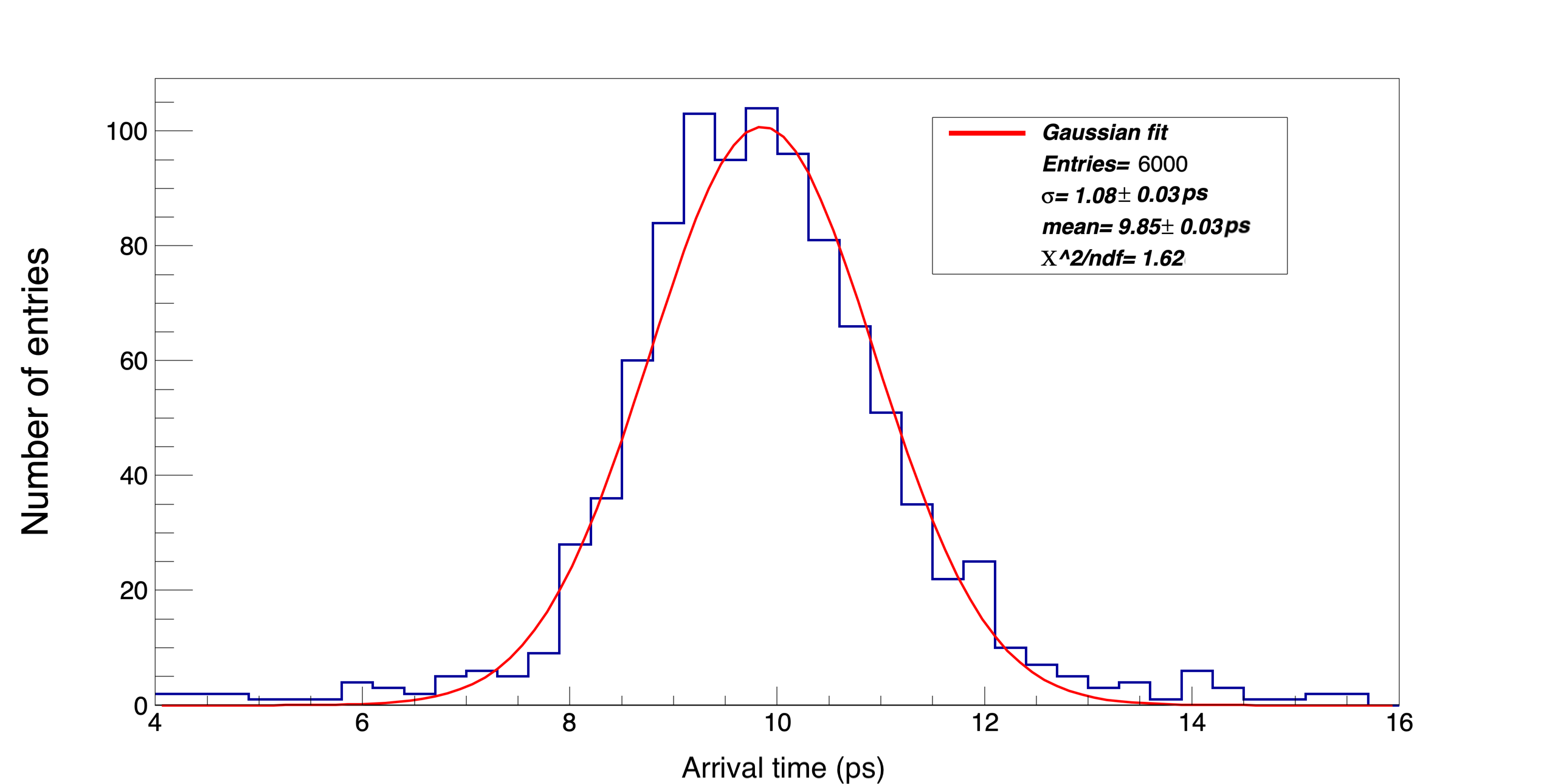}
\end{center}
\caption{Arrival time optical photons distribution produced by $\mu^-$ interacting to the first configuration.}
\label{muonmean}
\end{figure}

According to Table~\ref{source}, the Sr$^{90}$ source has similar energy than the rest of the sources, however, it has mass different from zero. As a consequence of these characteristics the $e^-$ is stopped at a certain depth, as it is shown in Figure~\ref{distance} for all configurations. It can be noted that the depth is the same, because the same material is used. For the hexagonal configurations, the $e^-$ is stopped at 10$\%$ of the width, therefore, very few photons are created compared with the other sources, which go through the plastic and therefore create photons on their way. The optical photons produced by the Sr$^{90}$ in the first configuration travel a greater optical path and therefore the mean arrival time is greater as it can been seen in Figure~\ref{mean}.  For the second configuration, the mean arrival time for the Sr$^{90}$  is consistent with the other sources, due to the location of the Scorer, so the optical path is almost the same. For the case of the square configuration, it can been seen that the mean arrival time is slightly bigger, for this case the e$^-$ is stopped at 73.3\% of the width, once again the optical path is greater, however, the difference with the optical path of optical photons produced by the other sources it is not that big compared to hexagonal configurations.\\
For the hexagonal configurations the $\mu^-$ source, that although it is a massive particle, has the smaller ITR value, this is due to the greater energy of the particle, which its interaction with the BC-404 scintillator produces much more optical photons and therefore the variation of the values around the mean (ITR) is smaller. To show this trend, as an example in Figure~\ref{muonmean} is shown the AT for the first configuration.  

\subsection{Intrinsic efficiency}
\begin{figure}[htbp]
\begin{center}
\includegraphics[width=0.8\textwidth]{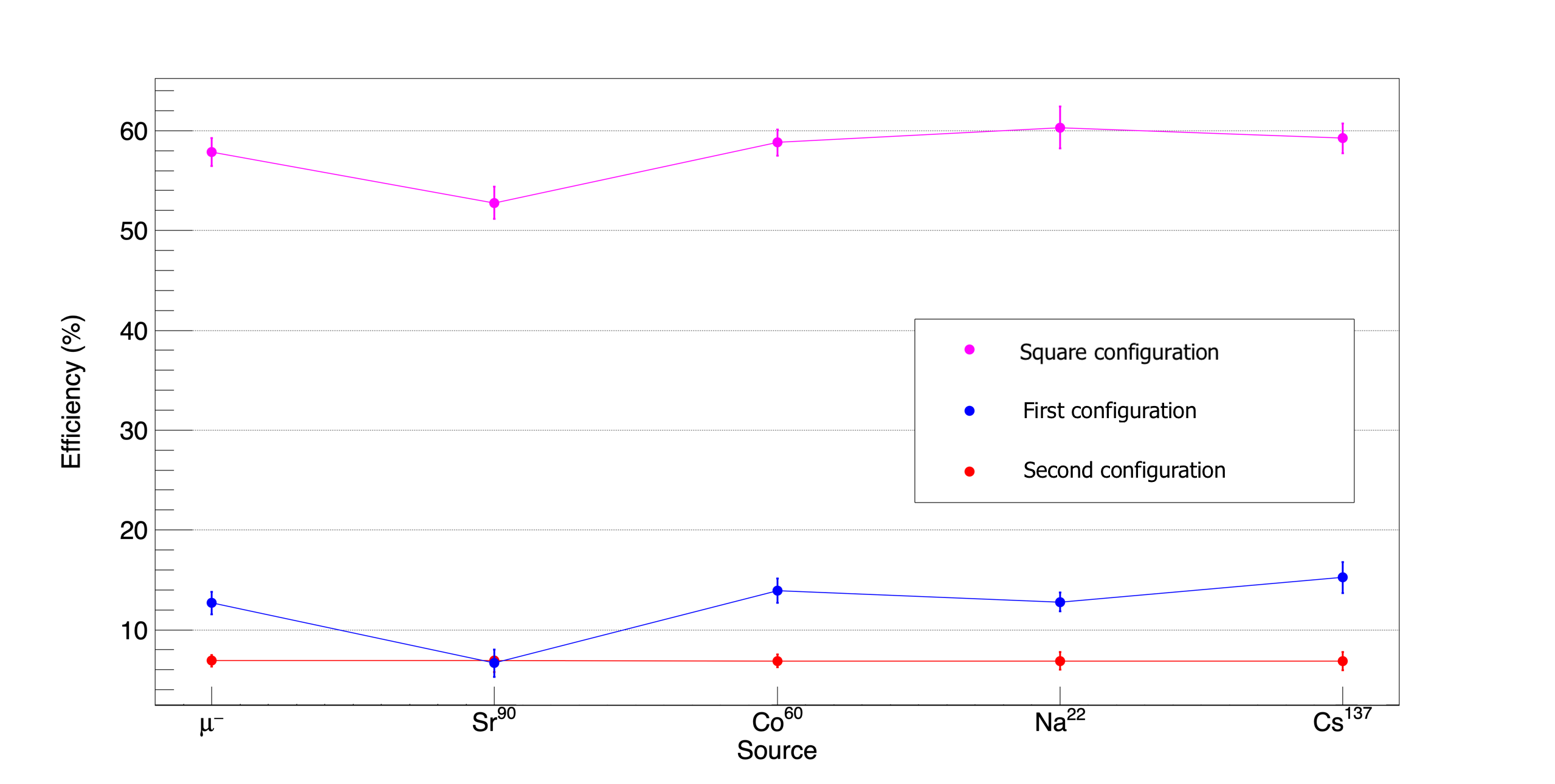}
\end{center}
\caption{Intrinsic efficiency values for all configurations.}
\label{eff}
\end{figure}

We define the intrinsic efficiency ($\epsilon_I$) of a detector as:
\begin{equation}
    \epsilon_I= \frac{N_d}{N_c}
\end{equation}
where $N_d$ is the number of optical photons detected on a Scorer and $N_c$ is the number of optical photos created. Analogously to ITR, $\epsilon_I$ depends of the size and geometry of the plastic scintillator and the location and size of the Scorer. In Figure~\ref{eff} is shown the $\epsilon_I$ values for all configurations and sources. Clearly, The square configuration has the best $\epsilon_I$ value and having better values for the other two configurations. This phenomenon is due to the optical path, that for the case of the square configuration, it is less than the hexagonal, then, the light attenuation becomes a relevant quantity for the hexagonal configuration due to its volume. Finally,  the location of the Scorer also influences as it can be seen for the first and second configuration. The effect of the depth for the Sr$^{90}$ source is reflected for second and square configurations. 

\subsection{Experimental measurement estimation}\label{charge_estimation}
\begin{figure}[htbp]
\begin{center}
\includegraphics[width=0.4\textwidth]{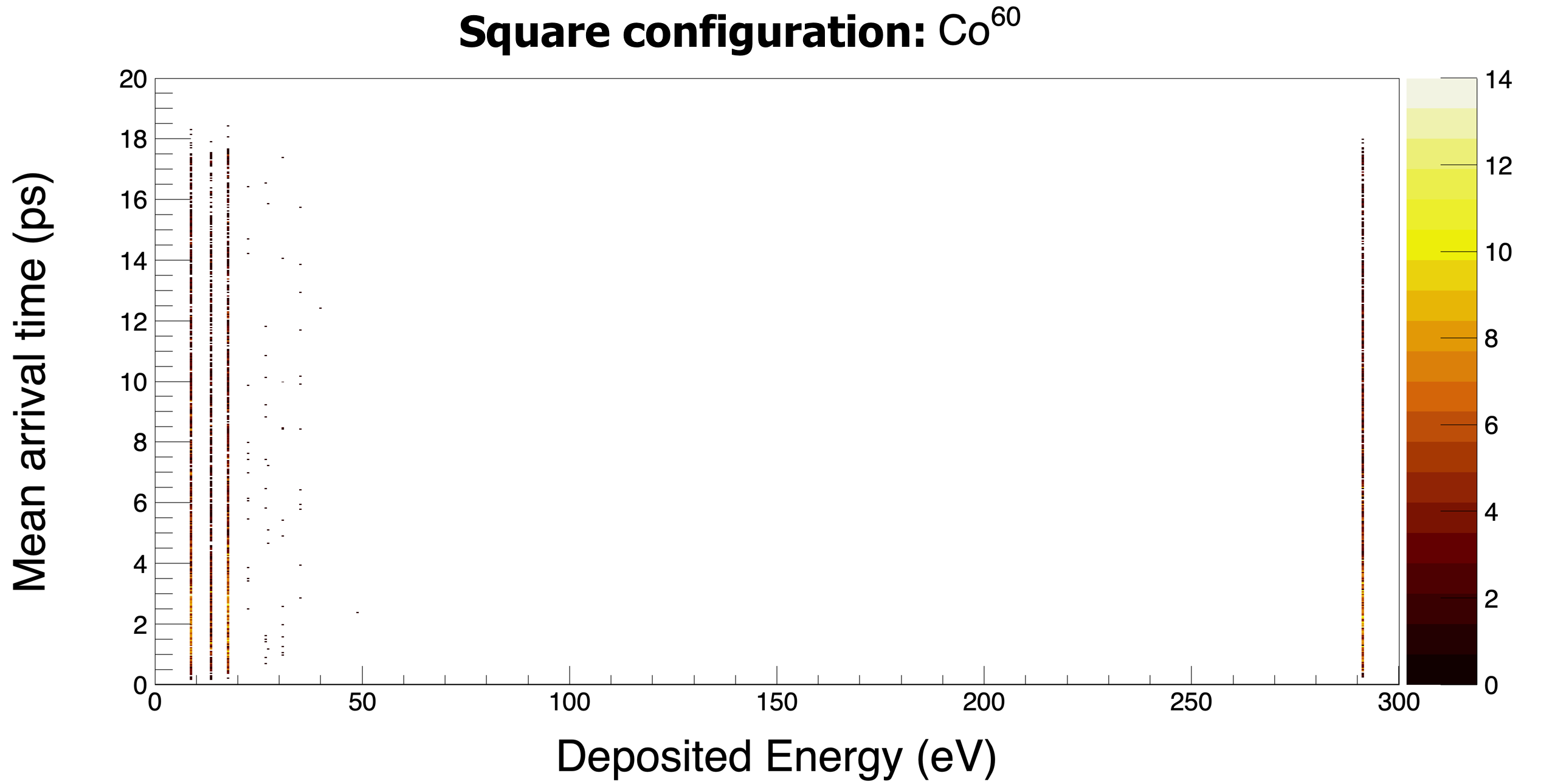}
\includegraphics[width=0.4\textwidth]{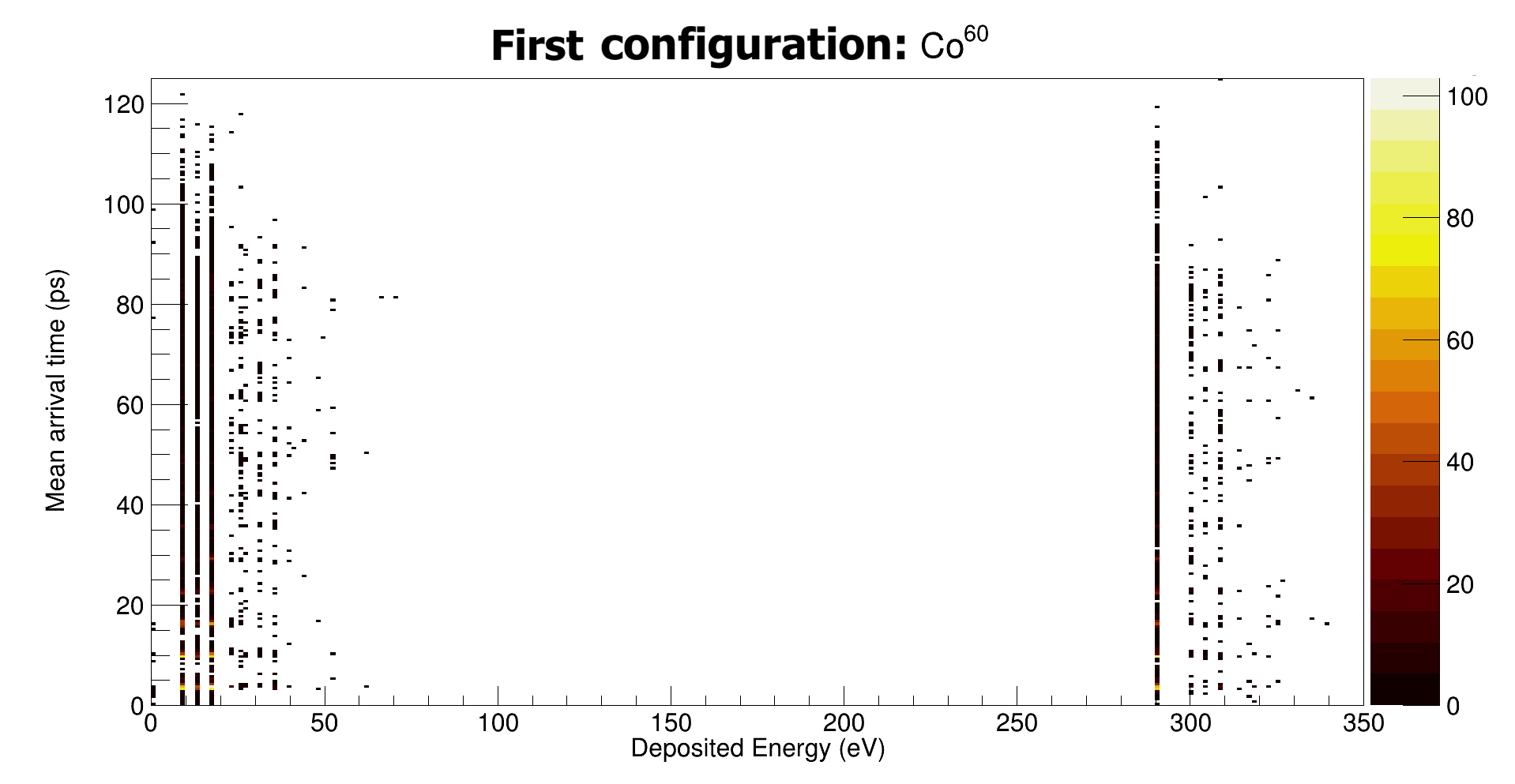}
\includegraphics[width=0.4\textwidth]{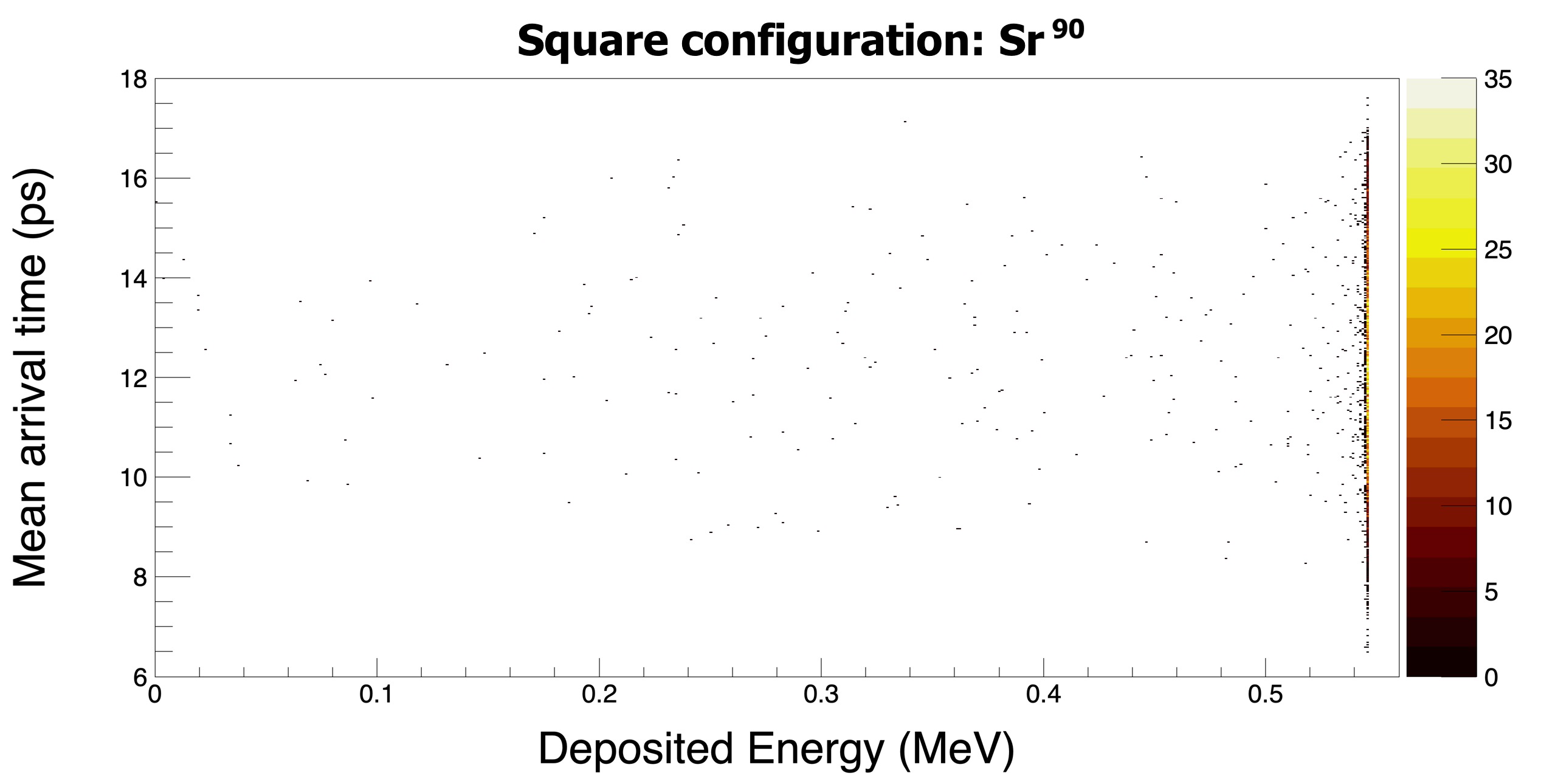}
\includegraphics[width=0.4\textwidth]{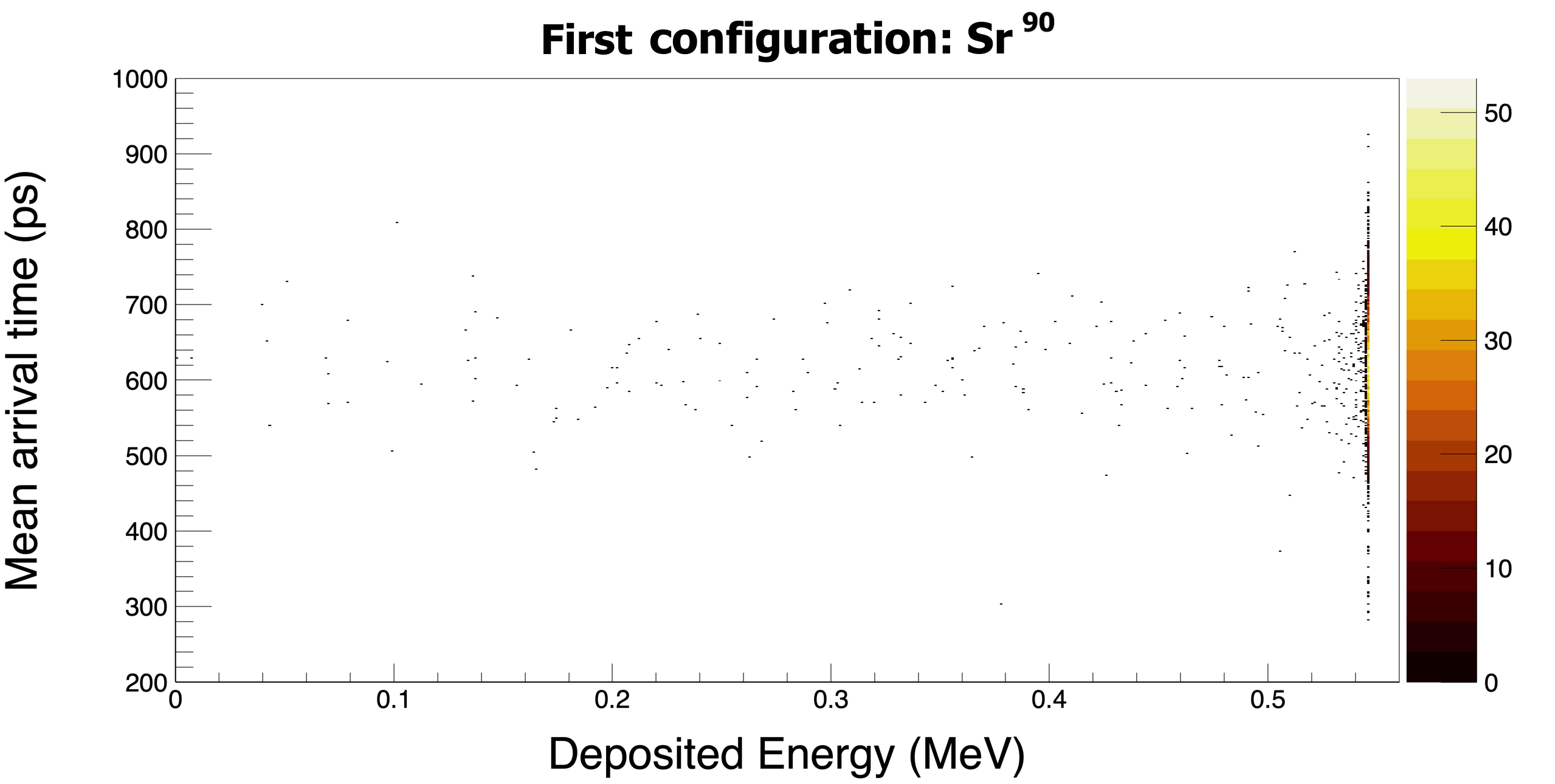}
\includegraphics[width=0.4\textwidth]{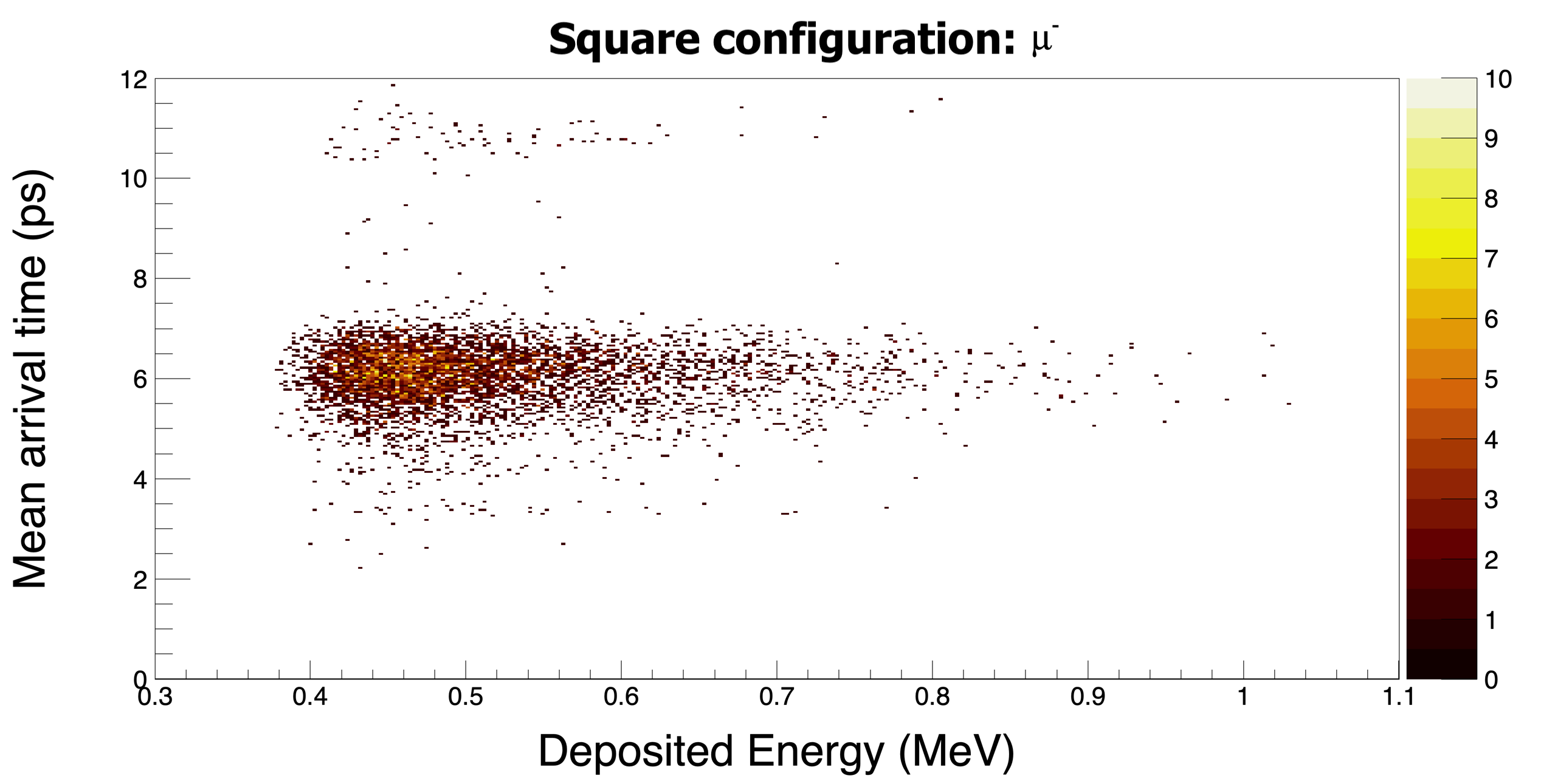}
\includegraphics[width=0.4\textwidth]{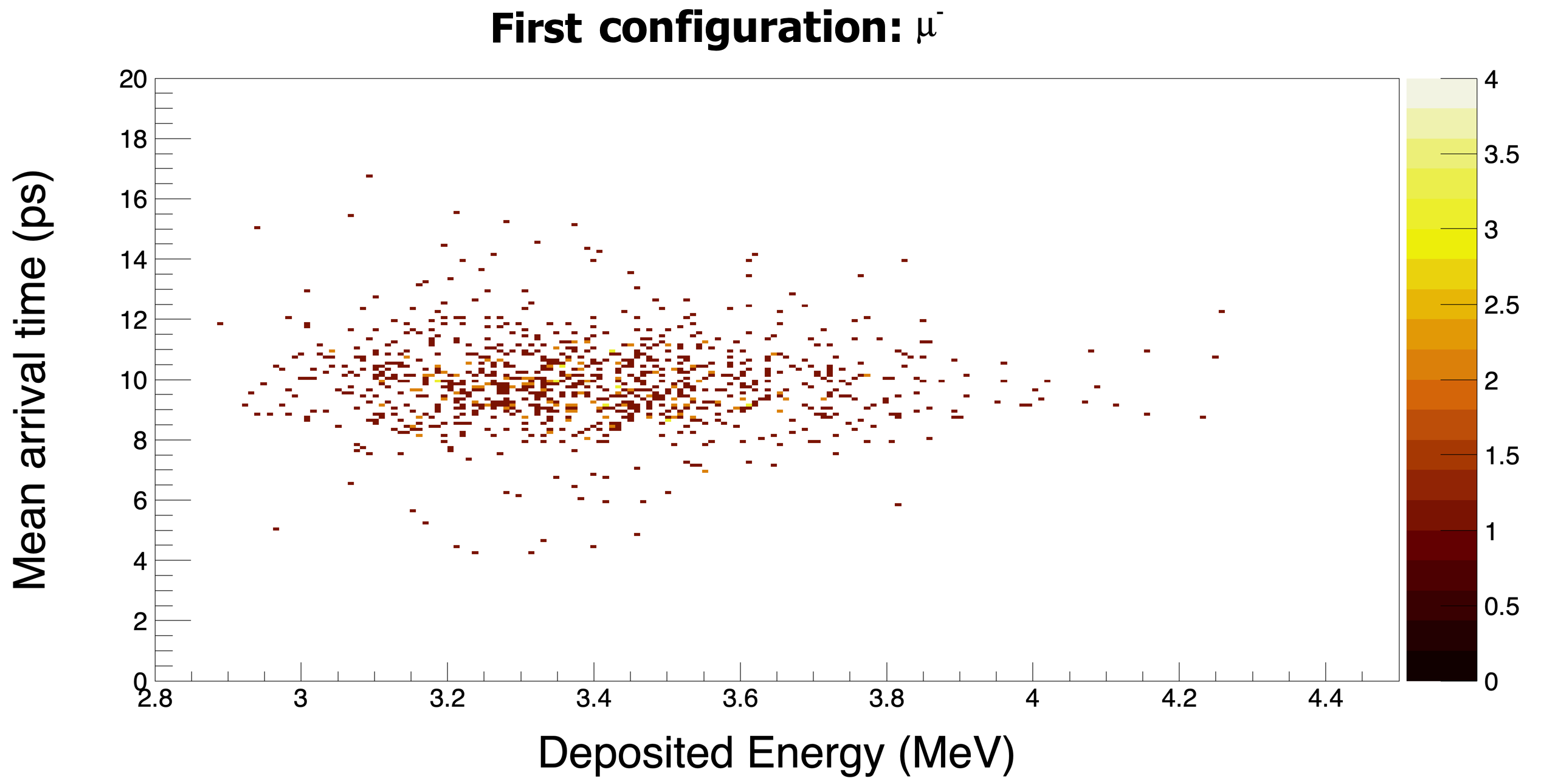}
\end{center}
\caption{Relation between mean arrival time and deposited energy for square configuration (left) and first configuration (right) for $Co^{60}$ (top row), $Sr^{90}$ (middle row) and $\mu^-$ (down row) sources.}
\label{EVsT}
\end{figure}

The deposited energy ($E_{dep}$) by the incident particle is related and increases or decreases with the charge deposited, which can be obtained from the pulse of the photo-senor~\cite{Zepeda_Fern_ndez_2020}. For the case where it is required to measure simultaneously the deposited charge and the response time of the detector, it is commonly used a Quad Digital Channel (QDC) and a Time to Digital Convert (TDC), respectively, to make a scatter plot. The AT is related to the TDC value, as it is shown in equation~\ref{Exp-Sim}. Then, the relation of AT and $E_{dep}$ is related to the values of TDC and QDC. As an example, in Figure~\ref{EVsT} are shown the AT and $E_{dep}$ relations for the Co$^{60}$, Sr$^{90}$ and $\mu^-$ sources, for the case of square and first configuration.

As it was mentioned above, the $\gamma$ particles hardly interact with plastic, for which an order of eV in $E_{dep}$ is obtained, also due to the nature of the particle, a quantified $E_{dep}$ is obtained. Also, it can be seen the effect that occurs with the electron, which does not pass through the plastic scintillator and therefore deposits all its energy. Finally, the $\mu^-$ source deposits more energy in hexagonal configurations than in square configuration, due to the width of the hexagon, in which, it interacts more.  Similar relations to Co$^{60}$ were obtained for the Na$^{22}$ and Cs$^{137}$ sources. For the second configuration we also obtained similar relations, having AT values greater than in the first configuration.\\



\section{Conclusions and discussion}
In this work we have shown that ITR and, therefore, the TR are not constant, that is, they depend of the energy, type of particle, the interaction point, the location of the Scorer (also the number of them~\cite{ALVARADO2020163150}) and the size and geometry of the plastic scintillator. The BC-404 plastic scintillator of small size exhibits an ITR close to 2~ps, which is also independent of the number of scorers~\cite{Kado_2021}. While for big size, the ITR and TR vary and in particular increase compare with a small size.\\
Another characteristic for a scintillator detector that should also be considered is the AT.  This quantity also depends on the features mentioned above. While the ITR can be similar or consistent for the kit source, the AT varies for Sr$^{90}$, due to the energy of the $e^-$, which does not pass through the material, therefore, this source is detected in a longer time than the rest of the kit sources.\\
Finally, the last characteristic that we propose for a detector is the intrinsic efficiency, which, allows to know the amount of optical photons detected by the configuration and geometry of a detector and the incident particle.\\
The low ITR values for the square configuration can be complicated to measure with conventional equipment, however, recently a TDC has been developed, whose RMS is of 2.2~ps~\cite{picoTDC}. The ITR value for the hexagonal configurations is more accessible to observe in a laboratory. The results obtained in this work are consistent with the results obtained in~\cite{ALVARADO2020163150}, which the incident particles were $\pi^+$ of 0.5~GeV and from which a ITR of 45~ps was obtained.\\
This simulations were made in order that they can be carried out in the laboratory, when using the  kit sources. By using a TDC and a QDC, the shape of the Figure~\ref{EVsT} can be recreated. In fact, the intrinsic efficiency also can be calculated. 


\end{document}